\documentclass[11pt]{article}

\usepackage{fullpage}
\usepackage{setspace}
\usepackage{parskip}
\usepackage{titlesec}
\usepackage[section]{placeins}
\usepackage{xcolor}
\usepackage{breakcites}
\usepackage{lineno}
\usepackage{booktabs}

\usepackage{amsmath}
\usepackage{amsfonts}
\usepackage{graphicx,psfrag,epsf}
\usepackage{enumerate}
\usepackage{natbib}
\usepackage{apalike}
\usepackage{url} 

\usepackage{glossaries}
\usepackage{glossary-mcols}
\makeglossaries

\usepackage[utf8]{inputenc}
\newcommand{\argmax}{\operatorname*{\arg\max}}

\newcommand{\NEG}[1]{\mbox{-}#1}

\usepackage{multicol}
\usepackage{multirow}
\usepackage{tabularx}
\usepackage{booktabs}
\usepackage{algpseudocode}
\usepackage{subcaption}
\newtheorem{prop}{Proposition}
\usepackage{color}
\usepackage{caption}
\usepackage{float}
\usepackage{tikz}
\usetikzlibrary{arrows,automata,backgrounds,fit,patterns,petri,shadows,shapes}
\pgfdeclarepatternformonly{stripes}
{\pgfpointorigin}{\pgfpoint{0.25cm}{0.25cm}}
{\pgfpoint{0.25cm}{0.25cm}}
{
    \pgfpathmoveto{\pgfpoint{0cm}{0cm}}
    \pgfpathlineto{\pgfpoint{0.25cm}{0.25cm}}
    \pgfpathlineto{\pgfpoint{0.25cm}{0.12cm}}
    \pgfpathlineto{\pgfpoint{0.12cm}{0cm}}
    \pgfpathclose%
    \pgfusepath{fill}
    \pgfpathmoveto{\pgfpoint{0cm}{0.12cm}}
    \pgfpathlineto{\pgfpoint{0cm}{0.25cm}}
    \pgfpathlineto{\pgfpoint{0.12cm}{0.25cm}}
    \pgfpathclose%
    \pgfusepath{fill}
}
\tikzstyle{decisionA} = [regular polygon, regular polygon sides = 4, thick, minimum size = 1.0cm, draw = black, fill = gray!40]
\tikzstyle{decisionD} = [regular polygon, regular polygon sides = 4, thick, minimum size = 1.0cm, draw = black]
\tikzstyle{utilityA} = [regular polygon, regular polygon sides = 6, thick, minimum size = 1.0cm, draw = black, fill = gray!20]
\tikzstyle{utilityD} = [regular polygon, regular polygon sides = 6, thick, minimum size = 1.0cm, draw = black]
\tikzstyle{chanceA} = [circle, thick, minimum size = 1.0cm, inner sep = 0.1pt, draw = black, fill = gray!40]
\tikzstyle{chanceD} = [circle, thick, minimum size = 1.0cm, inner sep = 0.1pt, draw = black]
\tikzstyle{chance} = [circle, thick, minimum size = 1.0cm, inner sep = 0.1pt, draw = black, pattern = stripes, pattern color = gray!40]
\tikzstyle{empty} = [circle, line width = 0pt, minimum size = 1cm, inner sep = 0.1pt]

\newcommand{\Acal}{\mathcal{A}}
\newcommand{\Dcal}{\mathcal{D}}

\newcommand{\Pbb}{\mathbb{P}}
\newcommand{\opt}{*}
\newcommand{\dd}{\mathop{}\! \mathrm{d}}

\usepackage{times}

\PassOptionsToPackage{hyphens}{url}
\usepackage[colorlinks = true,
            linkcolor = blue,
            urlcolor  = blue,
            citecolor = blue,
            anchorcolor = blue]{hyperref}
\usepackage{etoolbox}


\usepackage{natbib}

\renewenvironment{abstract}
  {{\bfseries\noindent{\abstractname}\par\nobreak}\footnotesize}
  {\bigskip}

\titlespacing{\section}{0pt}{*3}{*1}
\titlespacing{\subsection}{0pt}{*2}{*0.5}
\titlespacing{\subsubsection}{0pt}{*1.5}{0pt}

\usepackage{authblk}

\usepackage{graphicx}
\usepackage[space]{grffile}
\usepackage{latexsym}
\usepackage{textcomp}
\usepackage{longtable}
\usepackage{tabulary}
\usepackage{booktabs,array,multirow}
\usepackage{amsfonts,amsmath,amssymb}
\providecommand\citet{\cite}
\providecommand\citep{\cite}

\newif\iflatexml\latexmlfalse

\AtBeginDocument{\DeclareGraphicsExtensions{.pdf,.PDF,.eps,.EPS,.png,.PNG,.tif,.TIF,.jpg,.JPG,.jpeg,.JPEG}}

\usepackage[utf8]{inputenc}
\usepackage[english]{babel}

\usepackage{float}

\usepackage[margin=1.5in]{geometry}

\newcommand{\bc}{\begin{center}}
\newcommand{\ec}{\end{center}}
\newcommand{\Pb}{\hbox{{I}\kern-.1667em\hbox{P}}}
\newcommand{\Ex}{\hbox{{I}\kern-.1667em\hbox{E}}}
\newcommand{\Real}{\hbox{{I}\kern-.1667em\hbox{R}}}
\newcommand{\sReal}{\hbox{\scriptsize {I}\kern-.1667em\hbox{R}}}
\begin{document}

\title{Adversarial Risk Analysis (Overview)}

\author[1]{David Banks}
\author[2]{Victor Gallego}
\author[2]{Roi Naveiro}
\author[3,2]{David R\'ios Insua}%
\affil[1]{Duke University}
\affil[2]{Instituto de Ciencias Matematicas, ICMAT-CSIC}%
\affil[3]{University of Shanghai for Science and Technology}

\vspace{-1em}

\newcommand{\be}{\begin{equation}}
\newcommand{\ee}{\end{equation}}

  \date{}

\begingroup
\let\center\flushleft
\let\endcenter\endflushleft
\maketitle
\endgroup

\selectlanguage{english}
\begin{abstract}
Adversarial risk analysis (ARA) is a relatively new area of research that informs decision-making when facing intelligent opponents and uncertain outcomes. 
It enables an analyst to express her Bayesian beliefs about an opponent's utilities, capabilities, probabilities and the type of strategic calculation that the 
opponent is using. 
Within that framework, the analyst then solves the problem from the perspective of the opponent while placing subjective probability distributions on all unknown quantities. 
This produces a distribution over the actions of the opponent that permits the analyst to maximize her expected utility.
This overview covers conceptual, modeling, computational and applied issues in ARA.
\end{abstract}%

\section{Introduction}
Adversarial risk analysis (ARA) guides decision-making when there are intelligent opponents and uncertain outcomes. 
It is a decision-theoretic alternative to classical game theory that uses Bayesian subjective distributions to model the goals, resources, beliefs,
and reasoning of the opponent. 
Within this framework, the analyst solves the problem from the perspective of her opponent while placing subjective probability distributions on all 
unknown quantities. 
This provides a distribution over the actions of the opponent that enables her to maximize her expected utility, accounting for the uncertainty she has about the opponent.

Game theory is the standard approach to adversarial reasoning, and it has been applied, among many other areas, in politics \citep{brams2011game}, biology \citep{hammerstein1994game}, economics \citep{samuelson2016game}, social sciences \citep{shubik1982game} and cybersecurity \citep{shiva2010game}.
The cornerstone of game theory is the Nash equilibrium, in which no opponent can improve their outcome by any unilateral action. Of course,
this methodology has been generalized and extended in many ways \citep{halpern2008beyond}.
Nonetheless, the fundamental premises of game theory have been criticized \citep{hargreaves2004game,Young2004}.
The main concerns are:
\begin{itemize}
\item The classical formulation generally assumes that all participants in the game have the same beliefs about the other players, and that
all players know those beliefs are known.  
This common knowledge assumption is frequently unrealistic.
For example, in a three-person auction, it is quite possible for players A and B to have different distributions for the value to player C of the item on offer and that
they will conceal that information.
\item The field of behavioral economics has repeatedly demonstrated that humans do not act as game theory would prescribe \citep{camerer2003strategizing, gintis09}, so
it is a poor predictor of real-world decisions.
\item Game theory generally finds pessimistic and expensive solutions; it assumes that the opponent will do the most damaging thing
possible, and thus the analyst must invest in expensive protection. But often opponents are not so ruthless or so calculating.
\item The game theory paradigm assumes that each opponent is hyperlogical and capable of infinite computation.  
Researchers have weakened these assumptions in many ways;
e.g., through bounded rationality \citep{simon1955behavioral}, 
prospect theory \citep{tversky1979prospect}, and computing constraints \citep{osborne1998games}. But the core idea remains problematic.
\item Although the judgments of a supported agent could be arguably well assessed, 
as explained in \cite{KEENEY}, knowledge about the other agents' judgments is less
precise, since it requires them to reveal beliefs and preferences.
Such disclosure is unrealistic in cybersecurity, counterterrorism, and other domains,
where information is concealed by adversaries.
\item In spite of this, uncertainty in the adversary judgments is not
frequently acknowledged
and game theoretic solutions are often sensitive to such inputs as
shown in examples in \cite{ekin2019augmented}.
\item Game solutions can have multiple equilibria with no clear criteria to
choose among them \citep{raiffa2002negotiation}. 
Opponent A may select  a strategy corresponding to one equilibrium,
while opponent B could play a strategy corresponding to a different equilibrium;
see \cite{cooper} for developments in coordination games.
\item Except for toy problems, computing the game theoretic solution is difficult, often to the point of impracticability; see \cite{Nisan2007}  for issues
in computational game theory.
\end{itemize}
None of these criticisms is a sockdologer argument, and game theorists have worked hard to shore up these deficiencies.  
But although the equilibrium perspective is attractive, it is also compelling to maximize one's expected utility as in ARA.

Indeed, the ARA perspective is actually very natural. 
One builds a model for the thinking of one's opponent, and then maximizes expected utility under that model.
For example, when someone asks the boss for a raise, the first step is to understand what the boss values and what criteria he
will use. 
If that understanding is correct, the employee has a reasonable chance of success; but if the model for the boss's thought process is
wrong, then the outcome will likely be disappointing.

One of the advantages of ARA is its ability to partition the uncertainty into three separate components \citep{merrick2011comparative}.
The first is {\em aleatory uncertainty}, which is the uncertainty in the outcome conditional on the choices of each the opponents; 
this is handled by conventional statistical risk analysis 
\cite{bedford2001probabilistic, cox2013improving}.
The second component is {\em epistemic uncertainty}.
Usually, one does not know an opponent's utility function, nor his assessment of the probability of outcomes conditional on
the decisions that are made, nor even his resources;   
but a Bayesian should be  comfortable making subjective probability assessments about each of these quantities. 
The third component is {\em concept uncertainty}.
This reflects the fact that the analyst does not know how her opponent is making his decision.
Perhaps he is a game theorist and seeks an equilibrium solution.
Or perhaps he randomizes, or 
follows some other protocol.

To make this a little more concrete, consider a sealed bid auction between Daphne and Apollo,
each of whom wants to own a first edition of the {\em Theory of Games and Economic Behavior}.
Daphne's aleatory uncertainty is the value she receives conditional on her bid and Apollo's.
If she has not been allowed to examine the book prior to the auction, then its condition is a 
random variable--perhaps it is old and torn, or perhaps it has marginalia written by John Nash,
and both circumstances affect its value.
Epistemic uncertainty arises because Daphne does not know the value of the book to Apollo,
nor what he thinks is the probability that he will win with a bid of $x$ dollars, nor how much money Apollo has. 
The concept uncertainty reflects the fact that Daphne does not know whether Apollo is determining his
bid using classical game theory \citep{milgrom1982theory}, or whether he is simply bidding some unknown fraction of his true
top-dollar value, or using some other principle.

As we shall see, ARA can lead to mathematical formulations whose complexity is comparable to that found in game theory. There is no easy way to reason about other people's strategic thinking in realistic situations. Nonetheless, the decision-theoretic approach sidesteps some of the common criticisms of 
game theory.

\section{Auctions}\label{sec:auction}

To illustrate how ARA works, we shall build out the discussion of auctions. It is convenient to assume that Daphne and Apollo have had the opportunity to examine the book before the sealed-bid auction, which simplifies things by eliminating the aleatory uncertainty.
And suppose Daphne has a subjective
distribution $F(x)$ about the probability she will win the auction with a 
bid of $x$.  
Then, if her true top-dollar value for the book is $x_0$, 
assuming risk neutrality, she maximizes
her expected utility by bidding
$$
x^* = \mbox{argmax}_{x \in \mathbb{R}^+} (x_0 - x) F(x).
$$ 
Her utility is her profit $x_0 - x$, and the probability of realizing that
profit is $F(x)$.

How does Daphne acquire her subjective probability $F(x)$ regarding her
probability of winning?
There are several ways, and these depend upon which solution concept she thinks
Apollo will use.
In the simplest, she models Apollo as non-strategic, and believes he will
bid some unknown fraction of his true, also unknown, value.
As a Bayesian, Daphne is comfortable placing a subjective distribution 
$G(p)$ on the fraction of the true value that Apollo will bid ($0 \leq p \leq 1$),
and she can place a subjective distribution $H(v)$ on the value of the book
to Apollo.
Then, simple calculation shows that
\begin{equation}\label{nonstrategic}
F(x) = \Pb[PV \leq x] = \int_0^{\infty} \int_0^{x/v} g(p) h(v) dp dv
\end{equation}
where $g(p)$ and $h(v)$ are the densities of $G(p)$ and $H(v)$, respectively.

A second common solution concept Apollo might use is the Bayes Nash equilibrium (BNE).
The BNE formulation makes a strong common knowledge (CK) assumption: both Apollo
and Daphne have distributions $H_A$ and $H_D$ for each other’s valuation, and each
knows both distributions and knows that the other knows them.
This leads to solving a system of first-order ordinary differential equations. 
For an asymmetric auction, when $H_A \neq H_D$, no general solution exists, 
although it is known that if $H_A$ and $H_D$ are
differentiable then a unique solution exists and it is also differentiable 
\citep{Lebrun:1999}.
Previous attempts at solutions are based on the backshooting algorithm 
\citep{Hubbard:EtAl:2013}, but
\citet{Fibich:Gavish:2011} have shown that all such solutions are inherently
unstable.
\citet{au2014topics} has an algorithm that succeeds, based on the limit of
discretized bids and points of indifference.

From an ARA perspective, the CK assumption can be replaced by
something more reasonable. 
Daphne has a subjective belief about the distribution
$H_A$ of Apollo's value, and she has a subjective belief about
$H_D$, the distribution she believes he thinks is the distribution of her value.
The $H_A$ and $H_D$ are epistemic uncertainties.

In this BNE framework, she thinks Apollo will solve a system of coupled equations:
\begin{eqnarray}
\mbox{argmax}_{d \in \Real^+} (D^* - d) F_A(d) \sim F_D \nonumber \\
\mbox{argmax}_{a \in \Real^+} (A^* - a) F_D(a) \sim F_A
\end{eqnarray}
where $D^* \sim H_D$, $A^* \sim H_A$, and $F_A$ and $F_D$ are the
distributions of Apollo's and Daphne's bids, respectively.
The equilibrium solution gives $F_A$, her best guess, under
the BNE solution concept, of the distribution for Apollo’s bid.
Now Daphne should step outside the BNE framework and solve
$
x^* = \argmax_{x \in \Real^+} (x_0 - x) F_A(x)
$
where $x_0$ is Daphne's true value for the book, which of course is known to her. 

For a two-person auction, this solution is the same as found from a BNE
analysis.
But with more than two persons, the solution will diverge: Daphne can 
believe their opponent will bid high, but also think that Apollo thinks
he will bid low.
This allows non-common knowledge, and to distinguish this case from the
BNE, we call it a "mirroring argument" since Daphne is trying to do the
analysis she thinks her opponents are doing.

A third important solution concept is level-$k$ thinking
\citep{stahl1995players}.
A level-0 bidder is non-strategic.
A level-1 bidder thinks the opponent is non-strategic, and optimizes
her bid against a model for his behavior (as previously discussed).
A level-2 bidder thinks her opponent is a level-1 thinker who is modeling
her as a level-0 thinker.
This can get complicated, but except in highly structured games such
as chess, empirical studies indicate that people rarely go beyond level-2
thinking~\citep{stahl1994experimental}.

To illustrate, suppose Daphne is a level-2 bidder.
She thinks Apollo is a level-1 bidder, who models her as a 
non-strategic bidder.
Assume she believes that Apollo thinks her true value for the 
book has the uniform distribution on $[\$100, \, \$200]$ and
that she bids a random fraction $p$ of her value, where
$F(p) = p^9$ of $0 \leq p  \leq 1$.
From (\ref{nonstrategic}), it follows that Apollo thinks her
bid will have the distribution
$$
F(d) = \left( \frac{d - 100}{200} \right)^9 \;\;\;\; 100 \leq d \leq 200.
$$

Apollo needs to find his best response, by making the bid $a^*$ such that
$$
a^* = \mbox{argmax}_{a \in \sReal^+} (A_0 - a) F(a)
$$
where $A_0$ is his true value for the book, which is a random
variable to Daphne since she does not know it.
Using basic calculus to find the maximum shows
$$
0 = \frac{d}{da} (A_0 - a) F(a) = \frac{9a^8}{200^9} (A_0 - a) -
\left( \frac{a}{200}\right)^9
$$
so Apollo's bid should be 90\% of his true value.

Daphne does not know Apollo's value, but as a Bayesian, she has
a subjective distribution for it.
Suppose she thinks it is the triangular distribution on on
$[\$140, \, \$200]$ with a peak at \$170.
Then it follows that she thinks his bid will have the triangular
distribution on $[\$126, \, \$180]$ with a peak at \$153.
If her true value for the book is \$175, then her expected 
profit is maximized with a bid of \$161.67, which 
completes the level-2 analysis.

\section{ARA in general}

In ARA one takes the side of one agent, using only her beliefs and
knowledge, rather than assuming CK and 
trying to solve all of the agents' problems simultaneously.
The selected agent must have
(1) a subjective probability about the actions of each opponent,
(2) subjective conditional probabilities about the outcome for every
set of possible choices, and
(3) accurate knowledge of her own utility function.
Obviously, in practice a person will only have approximate
knowledge of these quantities, but that is sufficient for many
applications.

Thus Daphne believes Apollo has probability $\pi_D(a)$ of choosing action
$a \in {\cal A}$.
She has a subjective probability $p_D(s \,|\, d, a)$ for each
possible outcome $s \in {\cal S}$ given every choice
$(d, a) \in {\cal D} \times {\cal A}$, where ${\cal D}$ is
Daphne's set of possible actions.
And she knows her own utility $u_D(d, a, s)$
for each combination of outcome and pair of choices.

Daphne maximizes her expected utility by choosing
the action $d^*$ such that
\begin{eqnarray*}
d^* &=& \mbox{argmax}_{d \in {\cal D}} \, \Ex_{\pi_D, p_D} [ u_D (d, A, S)]\\
&=&  \mbox{argmax}_{d \in {\cal D}} \, \int_{s \in {\cal S}} \int_{a
  \in {\cal A}} \, u_D(d, a, s) p_D(s \,|\, d, a)  \pi_D(a) \, da \, ds 
\end{eqnarray*}
where $A$ is the random action chosen by Apollo and $S$ is the random
outcome that results from choosing $A$ and $d$.

In practice, the most difficult quantity to obtain is $\pi_D(a)$.
The $p_D(s \, | \, d, a)$ is found by conventional risk
analysis and $u_D(d, a, s)$ is a personal utility.
Risk analysis and utility self-assessment are not easy, but both are
mature fields and researchers know how to proceed.

Previously, we laid out ARA methods for obtaining $\pi_D(a)$ in 
auctions for the cases of the
the non-strategic opponent, the Nash equilibrium
seeking opponent, the opponent whose analysis mirrors that of the
decision-maker, and the opponent who is a level-$k$ thinker.
Implementing these approaches imposes different cognitive
loads upon the analyst.

Table \ref{cognitiveload} shows how the cognitive load depends upon
the kind of ARA.
Each row corresponds to a different level of reasoning
in level-$k$ thinking.
It displays the quantities that Daphne must
assess in order to implement a level-$k$ analysis.
Row 0 corresponds to the utilities and beliefs of Daphne and
Apollo, as perceived by themselves.
Subsequently, row $k$ contains the additional utilities and 
probabilities that Daphne would have to assess in order
to perform a level-$k$ analysis. 
The upper case characters in rows 1 and greater indicate that these
quantities are all random variables to Daphne.


\begin{table}
\centering
\resizebox{0.7\columnwidth}{!}{
\begin{tabular}{@{}ccccccc@{}}
\toprule
& \textbf{1} & \textbf{2}               & \textbf{3} & \textbf{4} & \textbf{5} & \textbf{6} \\ \midrule
\textbf{0} &$u_A$      & $p_A(\cdot \vert d, a)$ &   $\pi_A(d)$ &  $u_D$          &    $p_D(\cdot \vert d,a)$        &     $\pi_D(a)$       \\
\textbf{1} & $U^1_A$    & $P^1_A(\cdot \vert d, a)$ &  $\Pi^1_A(d)$ &  $U^1_D$           &   $P^1_D(\cdot \vert d,a)$         &  $\Pi^1_D(a)$          \\
\textbf{2} & $U^2_A$    & $P^2_A(\cdot \vert d, a)$ &  $\Pi^2_A(d)$ &  $U^2_D$          &   $P^2_D(\cdot \vert d,a)$         &  $\Pi^2_D(a)$          \\
\vdots & $\vdots$    &  $\vdots$  & $\vdots$   &       $\vdots$     &  $\vdots$          & $\vdots$           \\ \bottomrule
\end{tabular} 
}
\caption{Cognitive load for different kinds of ARA.}\label{cognitiveload}
\end{table}

The first column contains the utility functions she ascribes to Apollo.
The second column contains the conditional probabilities of the
outcome, given her choice and Apollo's, that she ascribes to Apollo.
The third column contains what she thinks is Apollo's 
distribution over her choice. 
The fourth column contains what Daphne believes are the utility
functions that Apollo ascribes to her.
The fifth column contains the probabilities of the outcome,
conditional on both her action and Apollo's, that she believes
Apollo ascribes to her.
The sixth column contains her opinion of what Apollo thinks
is her distribution for he will do.

In terms of the table, different solution concepts require
information in different cells:
\begin{itemize}
\item Traditional game theory requires cells (0,1), (0,2), (0,4),
  (0,5) and assumes that these are CK.
\item The non-strategic adversary analysis requires cells (0,4),
  (0,5) and (0,6), where the (0,6) cell is assessed from historical data
  and/or expert opinion.
\item When the adversary seeks a Nash equilibrium
  solution, the analysis requires cells (0,4), (0,5) and (1,1), (1,2), (1,4), and (1,5). 
  It uses these last four cells to infer cell (0,6). 
\item The level-$k$ adversary approach requires cells (0,4), (0,5) and:
\begin{itemize}
\item for a level-1 analysis, cells (1,1), (1,2) and (1,3) can produce (0,6);
\item for a level-2 analysis, cells (2,4), (2,5) and (2,6) produce
  (1,3), which, with (1,1), and (1,2) then produce (0,6);
\item and so forth for larger $k$.
\end{itemize}
\item The mirror equilibrium approach requires cells (0,4), (0,5) and 
uses a consistency condition between (1,1), (1,2), (1,3) and (1,4),
(1,5) and (1,6) to produce (0,6). 
\end{itemize}
The main message is that all of these methods entail significant
effort.

\section{Basic concepts in ARA}\label{KK}
In this section, we compare game-theoretic and ARA concepts in two template models. Their basic structures may be simplified or made more complex, by removing or adding nodes. Assume there is a defender ($D$) who chooses her decision $d \in \Dcal$ and an attacker ($A$) who chooses his attack $a \in \Acal$. In principle, the agents are assumed to act so as to maximize expected utility (or minimize expected loss)  \citep{french1986decision}.
Bi-agent influence diagrams (BAIDS) \citep{banks2015adversarial} describe the problems: the circular nodes represent uncertainties, the hexagonal show utilities, and square nodes indicate decisions. 
The arrows point to decision nodes (meaning that decisions are made given the values of  preceding nodes) or chance and utility nodes (these events or consequences are influenced by predecessors).   
Colored nodes indicate relevance to just one of the agents (white, defender; gray, attacker), striped nodes are relevant to both agents.
\subsection{Simultaneous defend-attack games}\label{sec:simultaneous}
\begin{figure}[htbp]
\centering
\begin{tikzpicture}[->,>=stealth',shorten >=1pt,auto,node distance=1.5cm,
                    semithick, scale=0.9, transform shape]

  \node[chance](D)   {$S $};
  \node[decisionD] (A) [left of=D]  {$D$};
  \node[decisionA] (B) [right of=D] {$A$};
  \node[utilityD]  (C) [below of=A] {$U_D$};
  \node[utilityA]  (E) [below of=B] {$U_A$};

  \path (A) edge    node {} (D)
            edge    node {} (C)
            edge    node {} (E)
        (B) edge    node {} (D)
            edge    node {} (E)
            edge    node {} (C)
        (D) edge    node {} (C)
            edge    node {} (E);
\end{tikzpicture}
 \caption{Basic two player simultaneous defend-attack game.}
  \label{PEDO}
\end{figure}
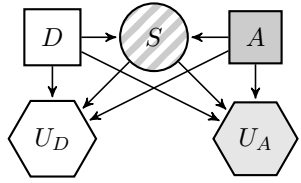

First consider simultaneous games---the agents decide their actions 
without knowing the action chosen by each other. 
Its template is shown in Fig.~\ref{PEDO}.

\noindent
The consequences for both agents depend upon the outcome of the
attack, $s \in S$. Each agent has their own
assessment on the probability of $s$,
which depends on $d$ and $a$, respectively called $p_D(s \vert d,a)$ and
$p_A(s  \vert d,a)$. Similarly, their utility functions are
$u_D(d, a, s)$ and $u_A(d, a, s)$.
Both agents $A$ and $D$ know the expected utility that a pair $(d,a)$ would provide them,
$ \psi_A (d,a) = \int u_A (d, a, s ) p_A(s | d,a) \dd s$,
and  $ \psi_D (d,a ) = \int u_D (d, a, s ) p_D(s | d,a) \dd s$.
A Nash equilibrium (NE) $(d^*,a^*)$ in this game satisfies 
$
\psi_D(d^*,a^*) \geq \psi_D(d,a^*) \,\,\, \forall \, d \in {\cal D}$
 and 
 $
\psi_A(d^*,a^*) \geq \psi_A(d^*,a) \,\,\, \forall \, a \in {\cal A} .
$

If utilities and probabilities are not CK, one may
model  the game as one with incomplete information \citep{harsanyi1967games} using the notion of types: each player will have a type known to him but not to the opponent, and thus is 
private information.
The type $\tau_i  \in T_i$ determines
the agent's utility $u_i(d, a, s ,\tau_i) $ and probability $p_i (s  | d, a, \tau_i)$,
$i \in \{A, D \}$.
Harsanyi proposes Bayes-Nash equilibria (BNE) as solution concept,
still under a strong CK assumption: the adversaries' beliefs about  
types are CK through a common prior $\pi(\tau_D, \tau_A)$ (moreover, 
the players' beliefs about other uncertainties in the problem are also
CK). Define strategy functions by associating a decision with each type,
$d:\tau_D \rightarrow d(\tau_D) \in {\cal D}, $ 
and $a: \tau_A
\rightarrow a(\tau_A) \in {\cal A} $.  
Agent $D$'s expected
utility associated with a pair of strategies $(d, a)$,
given her type $\tau_D \in T_D $, is
$$
\psi_D (d(\tau_D) ,a,\tau_D) =
  \int
  \int  u_D( d(\tau_D), a, s , \tau_D ) 
  p_D(s | d(\tau_D), a(\tau_A), \tau_D )  
  \pi(\tau_A | \tau_D ) \; \mathrm{d}\tau_A 
  \mathrm{d}s . 
$$
Similarly, we compute the attacker's expected utility $\psi_A (d,a(\tau_A),\tau_A)$.
Then, a BNE is a pair $(d^*,a^*)$ of strategy
functions satisfying
\begin{eqnarray*}
\psi_D(d^*(\tau_D),a^*, \tau_D)  &\geq \psi_D(d(\tau_D),a^*, \tau_D) , & \forall \, \tau_D  \, \\
\psi_A(d^*,a^*(\tau_A), \tau_A) &\geq \psi_A(d^*,a(\tau_A), \tau_A) , &  \forall \, \tau_A
\end{eqnarray*}
for every $d$ and every $ a$, respectively.

However, the common prior assumptions are still
unrealistic \citep{Antos10}, e.g.\ in security contexts.
We thus weaken them when supporting $D$, who 
should maximize her expected utility through 
\begin{equation}\label{kantorei}
d^* = \operatorname*{arg\,max}_{d \in {\cal D}} \,
       \int  \int  u_D (d, a, s ) p_D (s  \mid d,a) \pi_D (  a ) \dd s \dd a  =
       \operatorname*{arg\,max}_{d \in {\cal D}} \,
       \int  \psi_D (d, a) \pi_D (  a ) \dd a,
\end{equation}
where $\pi_D (  a )$ models her beliefs about 
the attacker's decision $a$. To assess such probability distribution, suppose $D$ thinks that $A$ maximizes expected utility,
so that he seeks\\ 
$a^* = \argmax_{a \in {\cal A}} 
     \int   \left[ \int  u_A (d, a, s ) 
 p_A (s  |  d,a) \dd s \right]  \pi_A ( d ) 
      \dd d$. 
She will typically be uncertain about $A$'s required inputs   $(u_A, p_A, \pi_A)$. 
If one models all information available to her about these elements through a subjective probability distribution $F \sim (U_A,P_A, \Pi_A)$,
mimicking ~\eqref{kantorei}, one propagates the uncertainty into
the distribution of
\begin{equation}\label{DURHAM}
    A \mid D  \sim  \argmax_{a \in {\cal A}} \,
    \int   \left[ 
    \int  U_A (d, a, s ) \;
    P_A (s  \mid d,a)  \dd s  \right]
  \,    \Pi_A ( D = d ) \dd d.
\end{equation}
Usually, to estimate $\pi_D(a)$ we use Monte Carlo simulation, drawing $K$ samples from $F$ $\left \lbrace u_A^k, p_A^k, \pi_A^k \right \rbrace_{k=1}^K \sim F$, finding for each of them
\begin{equation*}
    A^{*}_k =   \argmax_{a \in {\cal A}} \,
    \int   \left[ 
    \int  u^k_A (d, a, s ) \;
    p^k_A (s  \mid d, a)  \dd s  \right]
  \,    \pi^k_A (d ) \dd d.
\end{equation*}
and approximating $\pi_D(a)$ using the empirical frequencies
\begin{equation*}
    \widehat{\pi}_D (d) = \frac{\# \left \lbrace A^*_k = d \right \rbrace}{K}
\end{equation*}
%


$U_A$ and $P_A$ may be directly elicited from $D$.
However, eliciting $\Pi_A$ may require deeper analysis and level-$k$ thinking as she needs to model how $A$ analyzes $D$'s problem
(this is why we condition on the distribution of $D$ in \eqref{DURHAM}). This entails computing  
$ 
    D \mid A^1  \sim  \argmax_{d \in {\cal D}} \,
    \int 
    \int U_D (d, a, s ) \;
    P_D (s  \mid d,a)  \dd s \,\, 
    \Pi_{D} ( A^1 = a ) \dd a,
$
assuming she is able to assess: $(U_D, P_D)$, representing her knowledge about how $A$ estimates $u_D(d,a, s)$ and $p_D (s | d,a)$; and $\Pi_{D} ( A^1 )$, her beliefs about $A$'s estimate of the probability of the defender playing action $d$. $A^1$ is $A$'s random decision within $D$'s second level of recursive thinking.  Again, eliciting this last element may require further thinking from $D$, leading to a recursion of nested models, related to the level-$k$ thinking concept in \cite{stahl1994experimental}. 
This recursion might stop at a level in which $D$  
lacks the information necessary to assess the corresponding
distributions. At that point, she could assign a non-informative distribution \citep{french1986decision}. Further details are in \cite{riosrios}.

In  general, and particularly in simultaneous games, ARA and game theory use different conditions and assumptions, so it is natural that they lead to different solutions see, for instance, a cybersecurity example in \cite{rios2019adversarial}. As shown in \cite{ekin2019augmented}, ARA solutions tend to be more robust than NE. This is related with the fact that NE solutions are based on point estimates of the attacker's judgments, whereas ARA solutions take account of the uncertainty about such estimates,  leading to solutions that better reflect the analyst's beliefs.
Also, in the particular case of simultaneous games, BNE and ARA solutions do not necessarily coincide.

\subsection{Sequential defend-attack games} \label{sec:sequential}

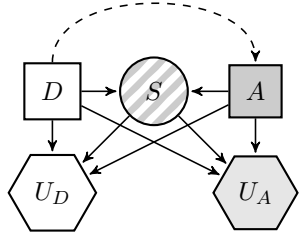
\begin{figure}[htbp]
\centering
\begin{tikzpicture}[->,>=stealth',shorten >=1pt,auto,node distance=1.5cm, scale=0.9, transform shape, semithick]

  \node[chance](D)   {$S  $};
  \node[decisionD] (A) [left of=D]  {$D$};
  \node[decisionA] (B) [right of=D] {$A$};
  \node[utilityD]  (C) [below of=A] {$U_D$};
  \node[utilityA]  (E) [below of=B] {$U_A$};

  \path (A) edge    node {} (D)
            edge    node {} (C)
            edge    node {} (E)
        (B) edge    node {} (D)
            edge    node {} (E)
            edge    node {} (C)
        (D) edge    node {} (C)
            edge    node {} (E)
        (A) edge[out=90, in=90, dashed] node {} (B);
\end{tikzpicture}
 \caption{Basic two player sequential defend-attack game}
 \label{fig:tpsdg}
\end{figure}

Consider now sequential games.
The defender chooses her decision $d$ and then $A$, after observing $d$, chooses his attack $a$. 
The BAID for this game is shown in Figure \ref{fig:tpsdg}.
The arc from $D$ to $A$ indicates that $D$'s choice is known to $A$.
This situation has been called the 
sequential Defend-Attack \citep{Brown:2006} game or Stackelberg game \citep{Gibbons:1992}.

The game-theoretic solution does not require $A$ to know
$D$'s judgments, as he observes her decision. 
However, $D$ must know those of $A$, which is the CK condition in this case. 
To solve, one computes both agents' expected utilities at node $S$: 
$\psi_A  (d, a)$ 
and
$\psi_D  (d, a)$. 
and finds
$a^\opt(d) = \argmax_{a \in \Acal}\, \psi_A(d,a)$,
$A$'s best response to $D$'s action $d$.
Then, $D$'s optimal action is
$d^\opt_\text{GT} = \argmax_{d \in \Dcal}\, \psi_D(d, a^\opt(d))$. The pair 
$\left( d^\opt_\text{GT},\, a^\opt(d^\opt_\text{GT}) \right)$ is a Nash
equilibrium and, indeed, a sub-game perfect equilibrium. 

The CK condition is weakened if we assume only partial information, leading to games under incomplete information \citep{harsanyi1967games}. In this case, to model $D$'s uncertainty about $A$'s elements, the attacker is assumed to belong to a certain type $\tau_A$, unknown to the defender. $A$'s optimal response to $d$ depends on his type, and is denoted as $a^*(d, \tau_A)$. If we model $D$'s uncertainty about $\tau_A$ through a prior $\pi(\tau_A)$, clearly, $D$'s optimal action is 
\begin{equation} \label{eq:BNE}
   \argmax_{d \in \Dcal} \int \psi_D[d, a^*(d, \tau_A)] \pi(\tau_A) \dd \tau_A. 
\end{equation}
%


The ARA approach to sequential games is different in its formulation. Given the lack of CK, $D$ is uncertain about the attacker's response to her action $d$ and models this uncertainty through the distribution $p_D(a \vert d)$.
Then, her expected utility would be
$
\psi_D(d) = \int \psi_D(d,a)\, p_D(a \vert d) \dd a $ 
with optimal decision $d^\opt_\text{ARA} = \arg\max_{d \in \Dcal} \psi_D(d)
$.

To elicit $p_D(a \vert d)$, $D$ benefits from modeling 
$A$'s problem. For this, she would use all information available about $u_A$ and $p_A$ to model her uncertainty about these elements  through a distribution $F = (U_A, P_A) $ over the space of utilities and probabilities. This induces a distribution over $A$'s expected utility, being his random expected utility is $\Psi _A (d,a) = \int U_A (d, a, s) P_A(s \vert d,a) \dd s$. 
Then, $D$ finds $p_D(a \vert d) = \Pbb_F \left[ a = \argmax_{x \in \Acal} \Psi_A(x,d) \right]$.  In general, one can use Monte Carlo (MC) simulation to approximate $p_D(a \vert d)$.

As in simultaneous games, NE and ARA solutions are different in this sequential setting, as they are based in different assumptions. However, BNE and ARA solutions coincide in this case, although their operational interpretations are quite different. 
As mentioned, computing the ARA solutions requires eliciting $p_D(a \vert d)$, which in turn requires modelling $D$'s uncertainty about $A$'s elements through a distribution $F = (U_A, P_A) $ over the space of utilities and probabilities. Without loss of generality, assume that both $U_A$ and $P_A$ are defined over a common probability space $(\Omega,{\cal A},{\cal P})$ with atomic elements $\omega \in \Omega$ \citep{Chung}. This induces a distribution over the Attacker's expected utility $\psi_A(d,a)$, being its random expected utility $\Psi _A^\omega  (d,a) = \int U_A^\omega  (d, a,s )
P_A^\omega (s \vert d,a) \dd s$.
In turn, this induces a random optimal alternative defined through 
$A^{*}(d)^{ \omega }= \argmax_{x \in \Acal} \Psi_A ^\omega (d, x)$.
Then, the defender would find
\begin{equation}\label{eq:ara-bne}
    p_D(a \vert d) = \Pbb_F \left[ A^* (d) = a  \right]=
    {\cal P} (\omega : A^* (d)^{\omega } = a).
\end{equation}
Now, $\omega$ and ${\cal P}$ could be reinterpreted, respectively, as the type $\tau_A$ and the common prior $\pi$ in Harsanyi's doctrine. Obviously, $P_A ^\omega $ and $ U_A ^\omega$ correspond to $A$'s probability and utility given his type, respectively; and $A^* (d)^{\omega } = a^*(d, \tau_A)$ in this new interpretation. With this in mind, the connection between BNE and ARA is straightforward. Notice that we can rewrite the BNE solution for the defender \eqref{eq:BNE} as 
\begin{equation*} 
   \argmax_{d \in \Dcal} \int \left[ \int \psi_D(d, a) \delta(a - a^*(d, \tau_A)) \dd a \right] \pi(\tau_A) \dd \tau_A,
\end{equation*}
where $\delta(\cdot)$ is the Dirac delta function. If we change the order of the integrals on $a$ and $\tau_A$ we get
\begin{equation}  \label{eq:ara-bne_1}
   \argmax_{d \in \Dcal} \int \psi_D(d, a) \left[ \int \delta(a - a^*(d, \tau_A))  \pi(\tau_A) \dd \tau_A \right] \dd a.
\end{equation}
Clearly,
\begin{equation}  \label{eq:ara-bne_2}
    \int \delta(a - a^*(d, \tau_A))  \pi(\tau_A) \dd \tau_A = {\cal P} (\tau_A : a^*(d, \tau_A) = a) = p_D(a \vert d),
\end{equation}
where the last step follows from \eqref{eq:ara-bne}. Inserting \eqref{eq:ara-bne_2} into \eqref{eq:ara-bne_1}, we recover the ARA formulation of the sequential game. Thus, in the sequential Defend-Attack game, one can interpret the ARA approach in terms of Harsanyi's, although the underlying principles are different. The take home message is that ARA provides a formal decision 
theoretic based approach to specify the types $\tau_A$ and the prior $\pi(\tau_A)$, thus facilitating the implementation of Bayesian games as we discuss in Section \ref{sec:more_than_one} in the general case of facing more than one opponent.

\section{Game Theory and ARA: Further relations}
ARA operationalizes the Bayesian approach to
games \citep{kadane1982subjective, raiffa1982art}, 
constructing a procedure to make probabilistic predictions
about the decisions of an opponent. This section details the relationship between ARA and core foundational concepts in game theory. 

We consider again two agents, a defender $D$ and attacker $A$. To simplify the discussion, assume there is certainty about the actions' consequences and the utilities the agents obtain (should the outcome conditional on the actions be random, one would operate with expected utilities after modeling uncertainty). Each agent has a finite set of possible actions, denoted $\Dcal$ and $\Acal$ respectively. We discuss the case in which both agents simultaneously implement their actions and obtain their respective utilities $u_i(d,a)$, $i \in \lbrace A, D \rbrace$, which they want to maximize. 
Assume one can build a distribution $\pi_D(a)$ that
expresses $D$'s beliefs about $A$'s decision. Then, the defender should solve
\begin{equation} \label{eq:ara_opt}
  d^* = \argmax_{d \in \Dcal}  \sum_{a \in \Acal} u_D(d,a) \pi_D(a).
\end{equation}
This is just the Bayesian approach to games, initiated by \cite{kadane1982subjective}, \cite{raiffa1982art} and \cite{raiffa2002negotiation} in non-constructive ways. It has been criticized by \cite{harsanyi82} and \cite{Myerson97}, among others. Before ARA, there was no formal methodology allowing the analyst to encode her subjective probabilities on the actions of the other agents. Of course, one could claim just deploying standard 
probability elicitation methods for such purpose \citep{OHagan2006},
however \cite{cookebook} demonstrate the benefits of the decomposition
adopted in ARA in adversarial elicitation.

Armed with this setting, we discuss the ARA perspective of several key concepts in game theory, beyond the relations with Nash and Bayes Nash equilibria  concepts
outlined above.


\paragraph{ARA solutions and dominance.}
A major tenet in game theory is that the actions proposed for the agents should be non-dominated. It is easy to show that ARA actions
are non-dominated under fairly general conditions
\citep{esteban20}.
\begin{prop}\label{prop:non_dom}
  If $\pi_D(a) > 0$ for every $a \in \Acal$, then the ARA action $d^*$ defined in \eqref{eq:ara_opt} is non-dominated.
\end{prop}


\paragraph{ARA solutions and iterative dominance}
Besides dominance, another key rationality concept in game theory is iterative dominance \citep{hargreaves2004game}. Note, however, that if the opponent's payoff is unknown to the defender, due to the absence of CK,
then the agent using ARA will not know which are the non-dominated actions of the rest. However, she can consider possible rankings of the corresponding random utilities and use those to her advantage. The strongest case is when the attacker always gets more utility with one of his actions than with any other, as perceived by the ARA analyst. 

Consider the defender's analysis of her opponent, then
\citep{esteban20}
\begin{prop}\label{prop:rnd_dom}
  Let $U_A(d,a)$ be the (random) utility that $A$ receives, according to $D$, when he implements his $a$ and $D$ plays $d$.
   Suppose that there are two actions $a$ and $a'$ for the attacker such that
  \begin{equation*}
    \max\,{\rm supp}[U_A(d, a)] < \min\,{\rm supp}[U_A(d, a')],
  \end{equation*}
  for every $d$. Then, $\pi_D(a) = 0$.
\end{prop}

As a consequence, one can perform an ARA analogue of iterative dominance as follows. First, eliminate the dominated actions of the defender by means of Proposition \ref{prop:non_dom}; then, eliminate actions of the attacker using Proposition \ref{prop:rnd_dom}.
Once completed, try again to eliminate actions from the defender, and repeat until no alternative can be eliminated.

ARA can mimic less demanding conditions than the dominance in Proposition \ref{prop:rnd_dom}, employing ideas from stochastic dominance \citep{Levy1998}, including \emph{state dominance}, \emph{first order stochastic dominance} or \emph{second order stochastic dominance}.

\paragraph{ARA and ficticious play}
One approach to forming $D$'s subjective probabilities about the actions of her opponent is to use data from previous interactions among the agents, when relevant history is available.  
For example, she might be a level-1 thinker who believes $A$ is non-strategic (a.k.a level-0 thinker). In that case, she can use a Dirichlet-multinomial model, e.g.~\cite{Gelman1995}, and the expected predictive distribution over $A$'s $i$-th action $a_i$ would be
\begin{equation}\label{eq:Dir_mult}
 \pi_D(a_i) = \frac{\alpha_i + x_i}{\sum_{j=1}^{|\Acal|} {(\alpha_j + x_j)}},
\end{equation}
where $|\Acal|$ is the cardinality of $\Acal$, $\alpha_i$ is the Dirichlet prior parameter over the probability of $a_i$ and $x_i$ is the number of times that $A$ previously implemented such action. This approach is related with the fictitious play methodology in game theory, summarized in \cite{menache2011network}. Under appropriate conditions \citep{brown1951iterative}, if
all players use fictitious play they converge to a NE as the number of plays grows to infinity. But ARA uses it for just one player, it incorporates prior information and we work with a small number of iterations. The approach is easily extended to cases in which agents have longer memories.

%

\paragraph{ARA and level-1 thinking}

ARA can aid in eliciting the analyst's beliefs about the other agent by thinking about the problem he would solve. This would be
\begin{equation*}
  a^* = \argmax_{a \in \Acal} \sum_{d \in \Dcal} u_A(d, a) \pi_A (d),
\end{equation*}
where $\pi_A (d)$ represents $A$'s beliefs about $D$'s actions. Since CK does not hold, $D$ does not know $u_A$ nor $\pi_A$, but she can  
describe that uncertainty with a random utility $U_A$ and a random probability $\Pi_A$, as previously discussed. This leads her to compute the random optimal action
\begin{equation}\label{eq:opt_adv}
  A^* = \argmax_{a \in {\cal A}} \sum_{d \in \Dcal}
  U_A(d, a) \Pi_A(d).
\end{equation}
Then, the first agent would find
\begin{equation*}
  \pi_D(a) = \Pr(A^* = a).
\end{equation*}
Typically, as mentioned in Section \ref{sec:simultaneous}, this would be estimated by simulation, sampling from $U_A$ and $\Pi_A$, computing the corresponding optimal action and estimating its probability through empirical frequencies.

\paragraph{ARA and level-$k$ thinking.}
Assessing the random distribution $\Pi_A(d)$ entails strategic thinking since $D$ needs to understand how $A$ will model her decision problem. This can lead to a hierarchy of decision-making problems as in Section \ref{sec:simultaneous}.
This analysis continues recursively, creating the hierarchy of nested decision models. Thus, $D$ selects her action based upon a chain of reasoning of the form ``I think that $A$ thinks that I think...'' which will go $k$ levels deep, depending on how sophisticated she believes her opponent is. This is the level-$k$ thinking approach in \cite{stahl1995players}, who make it operational by dynamic programming. However it seems more natural to stop the iteration when no more information is reasonably available, and then use a non-informative prior for the random probabilities and utilities.

\section{Further topics}
We cover now additional important topics in ARA. First of all, we focus on cases 
in which other rationalities rather than maximum expected utility is 
assumed on behalf of the adversaries.
In addition, we have just deal with interactions between two agents. Next, we discuss how the ARA methodology could be extended to multiagent setting. We also discuss more complex interactions between two agents than the ones introduced in Sections \ref{sec:simultaneous} and \ref{sec:sequential}. We end the section 
discussing computational aspects of ARA.

\subsection{Facing adversaries with other rationalities}
Throughout the discussion, we have emphasised that all agents seek to maximize expected utility. This is reasonable for the supported agent, as we have
her available for dialogue, but not so clear for the adversaries, who may 
be interested in hiding and concealing information. But it is possible to use prospect theory instead of utility functions, or other rationalities reflected 
in Section 3, and all the previous results follow in analogy. Another nice point is the analyst does not have to suppose her opponent is using a single solution concept. For example, she might believe her opponent has probability $1/3$ of seeking a BNE, $1/3$ of being a level-0 thinker, and $1/3$ of being a level-1 thinker. This creates a mixture model that expresses her full uncertainty, and ARA can be used to solve this more complex problem.

\subsection{Facing more than one adversary}\label{sec:more_than_one}

Consider a community of $n$ agents. The $i$-th agent has a finite set ${\cal A}_i = \{a_i^1,\ldots,a_i^{m_i}\}$ of possible actions, $i = 1,\ldots,n$. The agents simultaneously implement their actions $a = (a_1, \ldots, a_n)$ and obtain their respective utilities $u_i(a)$, $i = 1,\ldots,n$, which they want to maximize. 
%
Assume one can build a distribution $\pi_1(a_2,\ldots,a_n) = \pi_1(a_{\NEG{1}})$ that expresses the first agent's beliefs about the decisions made by the other agents. In ARA, the first agent should solve
\begin{equation*}
  a_1^* = \argmax_{a_1 \in {\cal A}_1}\,\sum_{j=1}^{N_1} {u_1(a_1,a_{\NEG{1}}^j)\,\pi_1(a_{\NEG{1}}^j)},
\end{equation*}
where $N_1$ is the number of possible combinations of opponents' actions. 
The reviewed approaches to form this probabilities in the two agent case, can be  extended to the multi-agent setting.

For instance, when performing fictitious play, as in the two agents case, agent one could use a Dirichlet-multinomial model where now the expected
predictive distribution over all agents' actions excluding the first one would be
\begin{equation*}
 \pi_1(a_{\NEG{1}}^k) = \frac{\alpha_k+x_k}{\sum_{j=1}^{N_1} {(\alpha_j + x_j)}},\quad k = 1,\ldots,N_1 
\end{equation*}
being $\alpha_k$ the Dirichlet prior parameter over the probability of $a_{\NEG{1}}^k$ and $x_k$ is the number of times that the other agents previously implemented such action. 

The level-1 thinking approach can also be extended to the multi-agent system. This entails that agent 1 thinks about the problem that each of the other agents would solve. For the $i$-th agent, it is 
\begin{equation*}
  a_i^* = \argmax_{a_i \in {\cal A}_i}\,\sum_{j=1}^{N_i} {u_i(a_i,a_{\NEG{i}}^j)\,\pi_i(a_{\NEG{i}}^j)},
\end{equation*}
where $N_i$ is the number of possible combinations of agent $i$ opponents' actions and $\pi_i(a_{\NEG{i}}^j)$ represents the beliefs of the $i$-th agent about the actions of the others. Since CK does not hold, the $i$-th agent does not know $u_i$ nor $\pi_i$, but she can describe that uncertainty with a random utility $U_i$ and a random probability $\Pi_i$, as previously discussed. This leads her to compute the random optimal action
\begin{equation*}
  A_i^* = \argmax_{x \in {\cal A}_i}\,\sum_{j=1}^{N_i} {U_i(a_i,a_{\NEG{i}}^j)\,\Pi_i(a_{\NEG{i}}^j)}.
\end{equation*}
Then, $\pi_1$ would be assessed using
\begin{equation*}
  \pi_1(a_i^j) = \Pr(A_i^* = a_i^j),\quad j = 1,\ldots,m_i.
\end{equation*}

Extending further steps in the level-$k$ thinking approach to the case of several opponents is theoretically straightforward, but comes at a higher computational cost. An example of the ARA framework deployed in reinforcement learning scenarios with multiple agents can be seen in \cite{gallego2019opponent}.
To mitigate the computational cost entailed by the large size of 
the sets of $N_1$ and $N_i$ above, we could assume conditional independence
of the actions of opponents.

\subsection{Facing more complex interactions}

\begin{figure}[htbp]
\begin{subfigure}[t]{0.30\textwidth}
\centering
\begin{tikzpicture}[->,>=stealth',shorten >=1pt,auto,node distance=1.7cm, semithick, scale=0.8, transform shape]

  \node[chance](D)   {$S $};
  \node[decisionD] (A) [left of=D]  {$D$};
  \node[decisionA] (B) [right of=D] {$A$};
  \node[utilityD]  (C) [below of=A] {$U_D$};
  \node[utilityA]  (E) [below of=B] {$U_A$};
  \node[empty]     (empty) [below of=D ]{};
  \node[empty]   (empty2) [below of=empty ]{};

  \path (A) edge    node {} (D)
            edge    node {} (C)
        (B) edge    node {} (D)
            edge    node {} (E)
        (D) edge    node {} (C)
            edge    node {} (E)
        (B) edge[out=90, in=90, dashed] node {} (A);
        
\end{tikzpicture}
\caption{Sequential attack-defend.}\label{fig:a-d}
\end{subfigure}
\hfill
\begin{subfigure}[t]{0.30\textwidth}
\centering
\begin{tikzpicture}[->,>=stealth',shorten >=1pt,auto,node distance=1.7cm, semithick, scale=0.8, transform shape]

  \node[chance](D)   {$S $};
  \node[decisionD] (A) [left of=D]  {$D$};
  \node[decisionA] (B) [right of=D] {$A$};
  \node[utilityD]  (C) [below of=A] {$U_D$};
  \node[utilityA]  (E) [below of=B] {$U_A$};
  \node[empty]     (empty) [below of=D ]{};
  \node[chanceD]   (V) [below of=empty ]{$V$};

  \path (A) edge    node {} (D)
            edge    node {} (C)
        (B) edge    node {} (D)
            edge    node {} (E)
        (D) edge    node {} (C)
            edge    node {} (E)
        (A) edge[out=90, in=90, dashed] node {} (B)
        (V) edge    node {} (D)
        (V) edge    node {} (E)
        (V) edge    node {} (C)
        (V) edge[bend left=90, dashed] node {} (A);
        
\end{tikzpicture}
\caption{ Sequential defend-attack with private information.}\label{fig:d-a_pi}
\end{subfigure}
\hfill
\begin{subfigure}[t]{0.30\textwidth}
\centering
\begin{tikzpicture}[->,>=stealth',shorten >=1pt,auto,node distance=1.7cm, semithick, scale=0.8, transform shape]

  \node[chance](D)   {$S $};
  \node[decisionD] (A) [left of=D]  {$D_1$};
  \node[decisionA] (B) [right of=D] {$A$};
  \node[utilityD]  (C) [below of=A] {$U_D$};
  \node[utilityA]  (E) [below of=B] {$U_A$};
  \node[empty]     (empty) [below of=D ] {};
  \node[decisionD]   (V) [below of=empty ]{$D_2$};

  \path (A) edge    node {} (D)
            edge    node {} (C)
        (B) edge    node {} (D)
            edge    node {} (E)
        (D) edge    node {} (C)
            edge    node {} (E)
        (A) edge[out=90, in=90, dashed] node {} (B)
        (D) edge[dashed]    node {} (V)
        (V) edge    node {} (E)
        (V) edge    node {} (C)
        (A) edge[bend right=90, dashed] node {} (V)
        (B) edge[bend left=90, dashed] node {} (V);
        
\end{tikzpicture}
\caption{Sequential defend-attack-defend.}\label{fig:d-a-d}
\end{subfigure}
\caption{Bi-agent influence diagrams for ARA templates.}
\end{figure}
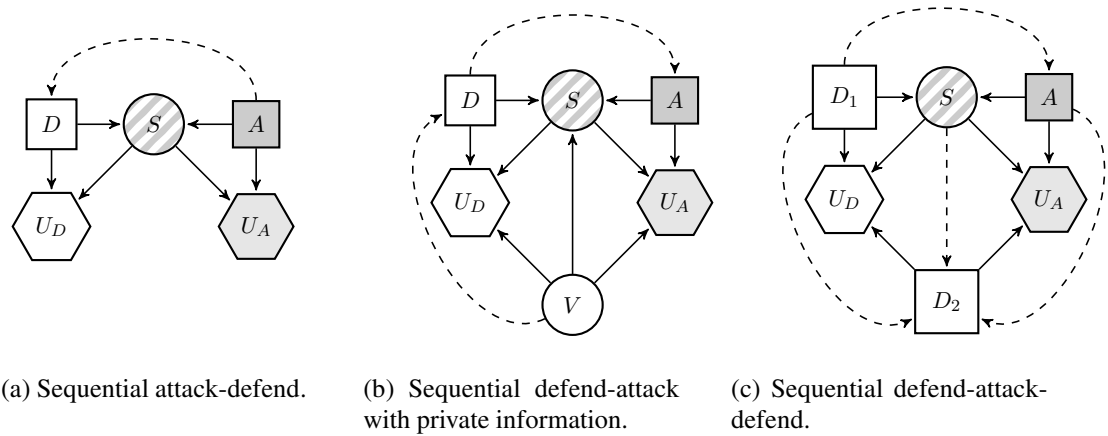

Previously, this paper introduced BAIDs for a simple simultaneous game and a simple sequential game.
Figure \ref{fig:a-d} reflects a template for the sequential defend-attack game
leading to the development of a contingency plan.
But ARA can treat more challenging situations. Figure \ref{fig:d-a_pi} generalizes the sequential defend-attack case of Figure \ref{fig:tpsdg} to situations in which the defender has private information $V$ not shared with the attacker. Arc $V-D$ shows that information is known by $D$ when she makes her decision; the lack of the arc $V-A$, shows that this information 
is not known by $A$ when making his decision. 
The uncertainty about the outcome $s$ depends on the actions by $A$ and $D$, as well as on $V$. 

Finally, Figure \ref{fig:d-a-d} depicts sequential defend-attack-defend games. In these, the defender moves first choosing $d_1$, then, the attacker chooses $a$ after having observed $d_1$, and finally the defender, knowing the outcome $s$, her previous decision $d_1$ and the attacker's $a$, 
chooses her final defense (mitigation) $d_2$.

Beyond these game structures, there can be more general and complex interactions between agents. Such BAIDS can be analyzed using methods
described in \cite{gonzalez2019adversarial}.


\subsection{An ARA computational pipeline}

Based on the BAID game templates, here is a general computational 
pipeline for ARA.\\
{\bf 1.  Modeling system threats.} Represent the attacker's
problem from the defender's perspective through 
an influence diagram (ID).
    The attacker's key features are his goals, knowledge and capabilities.
    Assessing these require determining
    the actions he may undertake and the utility that he perceives
    when performing them, given a defender's strategy. 
    The output is the set of attacker's decision nodes, together with the value node and arcs indicating how his utility depends on his decisions
    and those of the defender.
    Assessing the attacker's knowledge entails looking
    for relevant information that he may have when performing the attack, and his degree of knowledge about this information, as we do not assume CK. 
    This entails not only a modeling activity, but also a security 
    assessment of the system to determine which of its elements are accessible to the attacker.
    The outputs are the uncertainty nodes of the attacker ID,
    the arcs connecting them and  those ending in the
    decision nodes, indicating the information 
    available to $A$ when attacking. Finally, identifying his 
    capabilities requires determining which part of the defender
    problem the attacker has influence upon. This enables the attacker
    ID to connect with that of the defender.\\
    {\bf 2. Simulating attacks.} Based on step 1, a mechanism is required to simulate reasonable attacks.
     The state of the art solution assumes that the attacker seeks an NE, given strong CK hypothesis. The ARA methodology relaxes such assumptions,
        through a procedure to simulate adversarial decisions.
    Starting with the adversary model in step 1, the uncertainty about his probabilities and utilities is propagated to his problem and leads to the corresponding random optimal adversarial decisions which provide the required simulation. \\
    {\bf 3. Adopting defenses.} Augment the defender's
    problem by incorporating the attacker's problem produced in step 1. As output,
    generate a BAID reflecting the dual confrontation. 
    Finally, solve the defender's problem by maximizing her subjective expected utility, integrating out all attacker decisions, which are random from the defender's perspective given the lack of CK. In general, the corresponding integrals are
    approximated through MC, simulating attacks consistent with our
    knowledge about the attacker using the mechanism of step 2.
   
Thus, from a computational point of view, given a defense policy, one 
simulates from the attacker's problem to forecast the attacks and that enables the defender problem to find her optimal policies.
One can improve this general approach by taking advantage of the dynamic and informational structure of the BAID underlying the problem. One alternates between simulation and optimization as 
described in \cite{gonzalez2019adversarial}. But one still needs to simulate and optimize which can be computationally intensive. One possibility to alleviate this is to jointly the simulate and the optimize with the aid of the augmented simulation approach
\cite{ekin2019augmented}.

\section{Applications}

Section \ref{sec:auction} introduced basic concepts in ARA through an auction
example.
Further details may be seen in \cite{banks2015adversarial}.
Beyond that, applications in other areas abound. 
Many traditional applications of game theory in which CK conditions are debatable could be revisited from the ARA perspective. 
Important ARA applications are in defense, counter-terrorism, security, cybersecurity
risk analysis. 

Regarding defense, \cite{wang11} applies ARA to the problem of selecting a route through a network where an opponent chooses vertices for ambush. The methodology in that paper could be applied to convoy routing problems when there may be improvised explosive devices and imperfect information about their locations. \cite{sevillano12} uses ARA to support the owner of a ship in managing piracy risk; it models the situation using a defend-attack-defend game, in which CK is not assumed. \cite{Roponen1} considers ARA to enhance combat models
and \cite{Roponen2} uses it to protect from unmanned aerial vehicles.

The area of counterterrorism has also seen relevant applications of ARA. \cite{rios2012adversarial} provide general ARA models including defend-attack models, sequential defend-attack-defend models, and sequential defend-attack models with private information. These may be used as building blocks for more specific risk analysis of counterterrorism problems. For instance, \cite{Gil19} studies counter-terrorist online surveillance from an ARA perspective, focusing on the problem of monitoring a set of websites through classification of profiles  suspected of carrying out terrorist attacks. \cite{multimulti} deals with  critical infrastructure protection studying security resource allocation decision processes for an organization which faces multiple threats over multiple sites, specifically using a railways network.

Regarding urban security, applications in resource allocation have attracted interest.   
For example, \cite{gil16} uses ARA to allocate security resources to protect urban spaces, taking their spatial structure into account.
In cybersecurity, \cite{rios2019adversarial} provides a comprehensive framework for cybersecurity risk analysis, covering the presence of both intentional and non-intentional threats and the use of insurance as part of the security portfolio. \cite{naveiro2019adversarial} applies ARA to the emerging field of adversarial machine learning. In particular, it shows how to protect statistical classification systems from attackers trying to fool them by intentionally modifying input data.

Finally, it is worth mentioning ARA applications in social robotics, such as the one in \cite{esteban14}.
They illustrate how to use ARA to support the decision making of an autonomous agent that interacts with other agents and people in a competitive environment

\section{Conclusions}

This has been a brief overview on adversarial risk analysis,
its key concepts, methods and applications. 
But there remain many open topics. 

First, a promising line for future research is to extend ARA methodology to deal with
repeated play. The structure of these problems allows learning from past adversarial actions via Bayesian updating of the defender's beliefs about her opponent's unknowns. These models could be especially useful in games in which actions are taken in continuous time, also known as differential games \citep{dockner_jorgensen_long_sorger_2000}. In addition, interest in ARA in multi-agent reinforcement learning has recently risen \citep{gallego2019reinforcement}.  Research in multi-agent reinforcement learning has mainly focused in modeling the whole system through Markov games. However, the problem of supporting a single agent facing one or more opponents in a reinforcement learning setting is largely unexplored.

As we have seen, ARA relaxes crucial assumptions such as CK, thereby increasing realism. However, this comes at involved computations. Research on efficient approaches for ARA is crucial. For instance, exploring gradient-based techniques for sequential games is a fruitful line of research. \cite{naveiro2019gradient} provides an efficient solution method for sequential defend-attack games under the game-theoretic paradigm which could be extended to deal with the ARA setting. In addition, ARA entails simulating from the attacker's problem to forecast attacks and then optimizes for the defender to find her decision. This two-stage computation is demanding and single stage methods could reduce computation. Initial ideas based on augmented probability simulation are in \cite{ekin2019augmented}.

Finally, much research can be conducted in the applied side. For instance, regarding applications of ARA to adversarial machine learning problems, \cite{naveiro2019adversarial} aims at robustifying  classification systems against adversarial threats. Extensions of such techniques to regression or time series problems would be interesting. 
A broad overview on adversarial machine learning is in 
\cite{AMLARA}. Malware detection, fake news detection and autonomous vehicles security  are just a few examples of important societal applications of AML.

\section*{Funding Information}

{\label{808103}}
This work was partially supported by the National Science Foundation under Grant DMS-1638521 to the Statistical and Applied Mathematical Sciences Institute
and a BBVA Foundation project.
Roi Naveiro also acknowledges support of the Spanish Ministry for his grant FPU15-03636.
V.G. acknowledges also acknowledges support of the Spanish Ministry for his grant FPU16-05034.
David Rios Insua is grateful to the MTM2017-86875-C3-1-R AEI/ FEDER EU project,
the AXA-ICMAT Chair in Adversarial Risk Analysis and the EU's Horizon 2020 project 815003 Trustonomy.

\section*{Acknowledgments}

{\label{749861}}
 
We are grateful to discussions with J. Rios, A. Salo, J. Roponen, C. Joshi, F. Ruggeri, T. Ekin, P. Esteban, C. Gil, S. Liu and J. Ortega.

\selectlanguage{english}
\FloatBarrier

\bibliographystyle{apalike}
\bibliography{references}

\end{document}